\documentclass[fleqn,usenatbib]{mnras}

\usepackage{newtxtext,newtxmath}

\usepackage[T1]{fontenc}
\usepackage{ae,aecompl}

\usepackage{graphicx}	
\usepackage{amsmath}	
\usepackage{amssymb}	
\usepackage{placeins}

\title[Modelling the diffuse dust emission around Orion]{Modelling the diffuse dust emission around Orion}

\author[G. Saikia, P. Shalima \& R. Gogoi]{
Gautam Saikia,$^{1}$\thanks{E-mail: gautamsaikia91@gmail.com}
P. Shalima,$^{2}$\thanks{E-mail: shalima.p@gmail.com}
Rupjyoti Gogoi$^{1}$\thanks{E-mail: rupjyotigogoi@gmail.com}
\\
$^{1}$Department of Physics, Tezpur University, Napaam 784028, India\\
$^{2}$Regional Institute of Education Mysore, Mysuru, Karnataka 570001, India.\\
}

\date{Accepted XXX. Received YYY; in original form ZZZ}

\pubyear{2018}

\begin{document}
\label{firstpage}
\pagerange{\pageref{firstpage}--\pageref{lastpage}}
\maketitle

\begin{abstract}
We have studied the diffuse radiation in the surroundings of M42 using photometric data from the Galaxy Evolution Explorer (GALEX) in the far-ultraviolet (FUV) and infrared observations of the AKARI space telescope. The main source of the FUV diffuse emission is the starlight from the Trapezium stars scattered by dust in front of the nebula. We initially compare the diffuse FUV with the far-infrared (FIR) observations at the same locations. The FUV-IR correlations enable us to determine the type of dust contributing to this emission. We then use an existing model for studying the FUV dust scattering in Orion to check if it can be extended to regions away from the centre in a 10 deg radius. We obtain an albedo, $\alpha$ = 0.7 and scattering phase function asymmetry factor, g = 0.6 as the median values for our dust locations on different sides of the central Orion region. We find a uniform value of optical parameters across our sample of locations with the dust properties varying significantly from those at the centre of the nebula.
\end{abstract}




\section{Introduction}

The Orion Nebula (M42 or NGC 1976) is one of the most studied and nearest sites of active star formation [\cite{Bally+2008}] from the Milky Way (MW). It spans more than 700 deg$^2$ in the sky [\cite{Beitia-Antero+2017}] with emissions at different wavelengths. M42 is one of the brightest sources in the ultraviolet (UV) sky [\cite{Carruthers+1977}] and is one of the first objects targeted with new instruments for calibration purposes. The dust properties in Orion are known to be different from average Milky Way dust with the presence of much larger dust grains [\cite{Draine+2003}]. The brightness in Orion is attributed to the light from the Trapezium cluster of stars being forward scattered by the thin sheet of neutral hydrogen (HI), known as Orion's veil [\cite{Lockhart+1978}; \cite{O'dell+1993}], located $\sim$1 pc in front of the nebula. \cite{Abel+2004} state that the veil subtends an angle of at least 10' in the plane of the sky, which amounts to 1.5 pc at 500 pc. \cite{ODell+2001} has presented an extensive review of Orion nebula and related observations. \\

\cite{Murthy+2004} reported the first FUV diffuse observations in this region with data from the Far Ultraviolet Spectroscopic Explorer (FUSE) telescope. \cite{Shalima+2006} continued this work by modelling these observations to study the dust properties in the FUV and found that the albedo of these grains varies from 0.3 $\pm$ 0.1 at 912 \AA \hspace{0.1cm} to 0.5 $\pm$ 0.2 at 1020 \AA. The results of their proposed model were consistent with those of \cite{Draine+2003}. In the recent past, a lot of work has been done to study the dust properties in the central region of Orion, mainly concentrating on the $\sim$5' diameter optically bright region referred to as the Huygens region [\cite{Huygens+1659}] centered on the Trapezium stars, and the veil [\cite{Scandariato+2011}; \cite{Werf+2013}; \cite{Weilbacher+2015}; \cite{Schlafly+2015}; \cite{Beitia-Antero+2017}], but studies on the surroundings of Orion have been limited.\\

In this work, we present the findings after applying the dust model used successfully by \cite{Shalima+2006} at the Orion centre, to the surroundings of Orion in an unprecedented 10 deg radius. We have used archival data from the Galaxy Evolution Explorer (GALEX) [\cite{Martin+2005}] at FUV 1539 \AA \hspace{0.1cm} and from the AKARI space telescope [\cite{Murakami+2007}] archives at four far-IR wavelengths for this study. First we present the correlations between the GALEX FUV and the AKARI infrared observations. We then proceed to model the dust scattered FUV light in the Orion surroundings which will tell us about the scattering properties of the dust grains (albedo, cross-section and scattering phase function), and the nature of dust species as we move away from the centre of the nebula. We finally compare our results with previous work done for the Orion region and present our conclusions.\\

\section{Observations and data analysis} 

\subsection{The Data}

We have taken the location observed by \cite{Shalima+2006}, i.e. (l = 208.8, b = --19.3) as centre and looked for GALEX observations in a 10 deg radius. The GALEX mission performed surveys of the ultraviolet sky in two wavelength bands: FUV ($\lambda_{eff} \sim$ 1528 \AA, 1344--1786 \AA) and NUV ($\lambda_{eff} \sim$ 2310 \AA, 1771--2831 \AA) with different depth and coverage [\cite{Morrissey+2007}; \cite{Bianchi+2009}]. GALEX had a field of view $\approx$1.2$^0$ diameter, with a spatial resolution of $\approx$4.2$^{\prime\prime}$ (FUV) and $\approx$5.3$^{\prime\prime}$ (NUV) [\cite{Morrissey+2007}]. The two detectors provided simultaneous observations of the same field in two bands owing to the presence of a dichroic beam splitter [\cite{Bianchi+2014}]. The GALEX all sky imaging survey: AIS was completed in 2007 covering $\sim$ 26,000 deg$^2$ [\cite{Beitia-Antero+2017}].\\

\begin{figure*}
	\includegraphics[width=15cm]{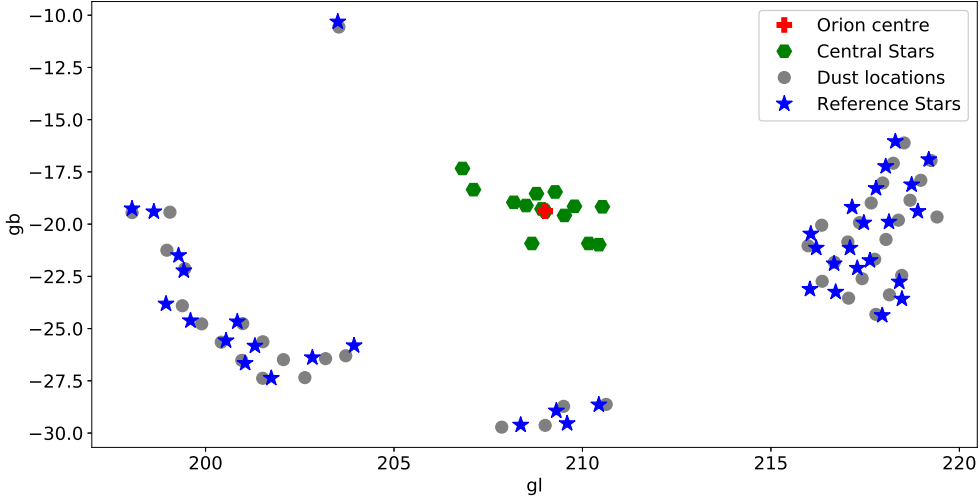}
    \caption{Our 42 locations around the Orion centre. We have distributed the locations into four groups based on their position with respect to the centre. The reference stars are used to get an estimate of the extinction and hydrogen column density near our dust locations with changing distance.}
    \label{our_locations}
\end{figure*}

Since Orion is very bright in the UV, GALEX didn't observe the central region of the nebula due to instrumental constraints and the observations start from an angular distance of 6.96 deg, which works to our advantage. We have an opportunity to apply a model working at the central region towards the outskirts of the nebula. We have taken 42 locations observed by GALEX in the FUV (1539 \AA) from the final data release of the spacecraft (GR6/GR7). We have 40 observations from GALEX AIS and 2 observations from GALEX Guest Investigator (GI) program. Using aperture photometry technique, we have calculated the flux at these locations and then converted them to intensities without convolving the images. The observed intensities have been corrected for airglow and background emission as described by \cite{Murthy+2014}. Our locations with respect to the Orion centre are shown in Figure \ref{our_locations}. The GALEX FUV details along with observed intensities are shown in Table \ref{galex_data}.\\

We have looked for infrared (IR) data at the same 42 locations in the AKARI legacy archive at four wavelength bands: 65 $\mu$m, 90 $\mu$m, 140 $\mu$m and 160 $\mu$m. These four wavelength bands were observed by the AKARI Far-Infrared Surveyor (FIS) [\cite{Kawada+2007}] which was the instrument chiefly intended to make an all-sky survey at far-infrared wavelengths [\cite{Doi+2015}; \cite{Takita+2015}]. The observation bands were named as: N60 (50--80 $\mu$m), WIDE-S (60--110 $\mu$m), WIDE-L (110--180 $\mu$m) and N160 (140--180 $\mu$m). The IR intensities $I_{65 \mu m}$, $I_{90 \mu m}$, $I_{140 \mu m}$ and $I_{160 \mu m}$ (with the subscript representing the wavelength in microns) observed by AKARI at our locations (calculated using aperture photometry) are shown in Table \ref{akari_data}.

\subsection{Correlation studies}

We have calculated the Spearman's rank correlations among the FUV (1539 \AA = 0.15 $\mu$m) and the four FIR wavelengths for which we have collected the archival data. The correlation value tells us both the strength and direction (positive or negative) of the monotonic relationship between two variables. It is non-parametric in the sense that it does not assume any model, like a straight line fit [\cite{Bevington+2003}]. The Spearman's rank correlation coefficient is calculated using the following relation:

$$ \rho = 1 - \frac{6 \Sigma d^2}{n^3  - n}$$
where, $\Sigma$ = sum, d = difference between two ranks, n = no. of pairs of data.\\

The observed correlations between the GALEX FUV data (Table \ref{galex_data}) and the AKARI FIR data (Table \ref{akari_data}) are shown in Table \ref{corr_table} and the corresponding graphs are shown in Figure \ref{corr_figs}. The probability or p-value is also shown which tells us how likely it is for the calculation to be a result of chance. A lower p-value signifies more reliability in the observed value of rank correlation coefficient.

\begin{table}
	\centering
	\caption{The rank correlation values among the FUV-FIR intensities for our 42 locations.}
	\label{corr_table}
	\begin{tabular}{ccc} 
		\hline
		FUV--FIR & Rank correlation ($\rho$) & p-value \\
		\hline
		 $I_{0.15 \mu m}$ $\sim$ $I_{65 \mu m}$  &  0.221  &   0.1642 \\
		$I_{0.15 \mu m}$ $\sim$ $I_{90 \mu m}$   &  0.545  &   0.0002 \\
		$I_{0.15 \mu m}$ $\sim$ $I_{140 \mu m}$  &  0.505  &   0.0007 \\
		$I_{0.15 \mu m}$ $\sim$ $I_{160 \mu m}$  &  0.401  &   0.0092 \\
		\hline
	\end{tabular}
\end{table}

\subsection{The FUV model}

We have used the model by \cite{Shalima+2006} and hence we try to constrain the albedo and the asymmetry factor of the dust grains in the region surrounding Orion. The model uses the Henyey--Greenstein scattering phase function [\cite{Henyey+1941}]:

$$ \phi(\theta) = \frac{(1-g^2)}{4\pi [1+g^2-2g cos(\theta)]^{\frac{3}{2}}}$$

where `g' is the phase function asymmetry factor and $\theta$ is the scattering angle. A value of `g' close to zero implies that the scattering is nearly isotropic while a value of g near 1 implies strongly forward scattering grains.\\

\begin{table*}
	\centering
	\caption{Properties of the 14 contributing central stars. The spectral type has been obtained from the SIMBAD astronomical database. The distance has been taken from the Hipparcos catalogue.}
	\label{stars}
	\begin{tabular}{lcccccr} 
		\hline
	HD number & l & b & Spectral type & Distance & Luminosity (1539 \AA) & N(H)\\
	&  ($^0$) & ($^0$) & & (pc) & (ph $cm^{-2}$ sec$^{-1}$ sr$^{-1}$ \AA$^{-1}$) & (cm$^{-2}$) \\
		\hline
	 037043   &   209.5221 & -19.5835   &       O9III  &  406.5041   &     07.453 $\times 10^9$   & 11.870$\times 10^{21}$\\
	 037022   &   209.0107 & -19.3840   &         O7V  &  450.4504   &     75.244 $\times 10^9$    &  342.245$\times 10^{21}$\\
	 036512   &  210.4356 & -20.9830    &     O9.7V  &  473.9337   &     09.443 $\times 10^8$   & 1.906$\times 10^{21}$\\
	 037017   &   208.1770 & -18.9571    &     B1.5V  &  373.1343    &    09.068 $\times 10^7$   &  6.508$\times 10^{21}$\\
	 037481   &   210.5290 & -19.1698    &     B2IV/V  &  480.7692    &    08.111 $\times 10^8$   & 8.952$\times 10^{21}$\\
	 037303   &      209.7887 & -19.1469   &        B2V  &  416.6667   &     02.842 $\times 10^8$   &  4.414$\times 10^{21}$\\
	 037061   &      208.9248 & -19.2736   &         O9V  &  361.0108  &      02.098 $\times 10^9$   &  230.939$\times 10^{21}$\\
	 037018   &      208.5036 & -19.1092   &        B1V  &  240.9639   &     02.352 $\times 10^9$   &   35.432$\times 10^{21}$\\
	 037356   &     208.7804 &  -18.5426   &        B3V  &  343.6426    &    02.753 $\times 10^8$   &   0.791$\times 10^{21}$\\
	 037526   &    209.2749 & -18.4597   &      B5V  &  332.2259    &    06.272 $\times 10^6$   &   1.715$\times 10^{21}$\\
	 036487   &    210.1646 & -20.9242   &       B6IV &   377.3585  &      04.621 $\times 10^6$   &   1.328$\times 10^{21}$\\ 
	 037468   &  206.8163 & -17.3360     &     O9.5V  &  352.1127    &    01.121 $\times 10^9$   &   2.317$\times 10^{21}$\\
	 037056   &    207.1087 & -18.3495   &       B8/9V  &  446.4286  &      05.391 $\times 10^6$   &   2.883$\times 10^{21}$\\
	 036120  &    208.6580 & -20.9250   &       B8V  &  390.6250     &   00.985 $\times 10^6$   &  0.527$\times 10^{21}$\\
			\hline
	\end{tabular}
\end{table*}

The Trapezium cluster of stars in Orion are the brightest source of radiation from the nebula as mentioned by \cite{Shalima+2006}. But since our region under consideration is much further away from the center of M42, we have taken into account the 14 brightest stars, in addition to the Trapezium stars as contributors of radiation at our locations as shown in Table \ref{stars}. The stellar luminosities shown in Table \ref{stars} have been calculated using data from the International Ultraviolet Explorer (IUE) archives at 1539 \AA. We have first used the all-sky 100 $\mu$m dust emission maps by \cite{Schlegel+1998} to extract the E(B-V) for our stars. We have then used the dust cross-section per hydrogen atom from \cite{Draine+2003} and an extinction R${_v}$ = 5.5 for Orion to calculate the total extinction A(V) and N(H) for each location:

$$\frac{A(V)}{E(B-V)} = R(V) = 5.5$$
$$\frac{\textit{N(H)}}{A(V)} = 1.8 \times 10^{21} \hspace{0.2cm} atoms \hspace{0.1cm} cm^{-2} \hspace{0.1cm} mag^{-1}$$

Since we do have much idea about the dust properties at our 42 locations, we have selected a set of reference stars as shown in Figure \ref{our_locations} from the Hipparcos catalogue [\cite{Perryman+1997}] close to our dust in order to get a better estimate of the extinction and distance to our locations. We use a CO map from Planck Telescope archive [\cite{Planck+2016}] to separate the molecular component, N(CO) from the total N(H). Now, the E(B-V) values of the stars varies differently with distance: very close to the centre the dust density is high, so we make a table of N(H) values for these stars and put them in bins of suitable size to allow us to compare with the values in the locations.\\

In our model, we only consider single-scattering due to the presence of Orion's Veil in front of the nebula which scatters the light from the 14 stars we have considered. We mainly put the N(H) values in front of the reference stars. The optical depth is given as, $\tau(\lambda) = N(H) \times \sigma$, where $\lambda$ = 1539 \AA \hspace{0.2cm} and $\sigma$ is the extinction cross-section. Since we have low values of $\tau$ ($\ll$1), single-scattering dominates. Hence, the reference stars in Figure \ref{our_locations} are used for distributing the material. If we see the column density remaining constant despite the distance, then we put the entire N(H)--N(CO) in front of the star. Otherwise we put two or three sheets of 0.1 pc so that we get the total column density. e.g. The star N(H)=4.6e21 is in front of the star. So we put 4.6e21/3e18/0.1 at 210pc and (6.69e21--4.6e21)/3e18/0.1 at 450pc (where 6.69e21 is the total N(H) for the dust location). Dividing the N(H) by the factor (3e18$\times$0.1) gives us the number density. Hence, if the column density is same with distance, then all the dust is in front of the closer star. Between that and the star which is away there is not much dust. Assuming the same albedo ($\alpha$), g, throughout and varying the distance and column densities, we try to get a good fit between the GALEX FUV observations and our model output. This will give us the 3D dust distribution (as seen in Figure \ref{dust_dist}).\\

\section{Results and Discussion}

\begin{figure*}
	\includegraphics[width=16cm, height=11cm]{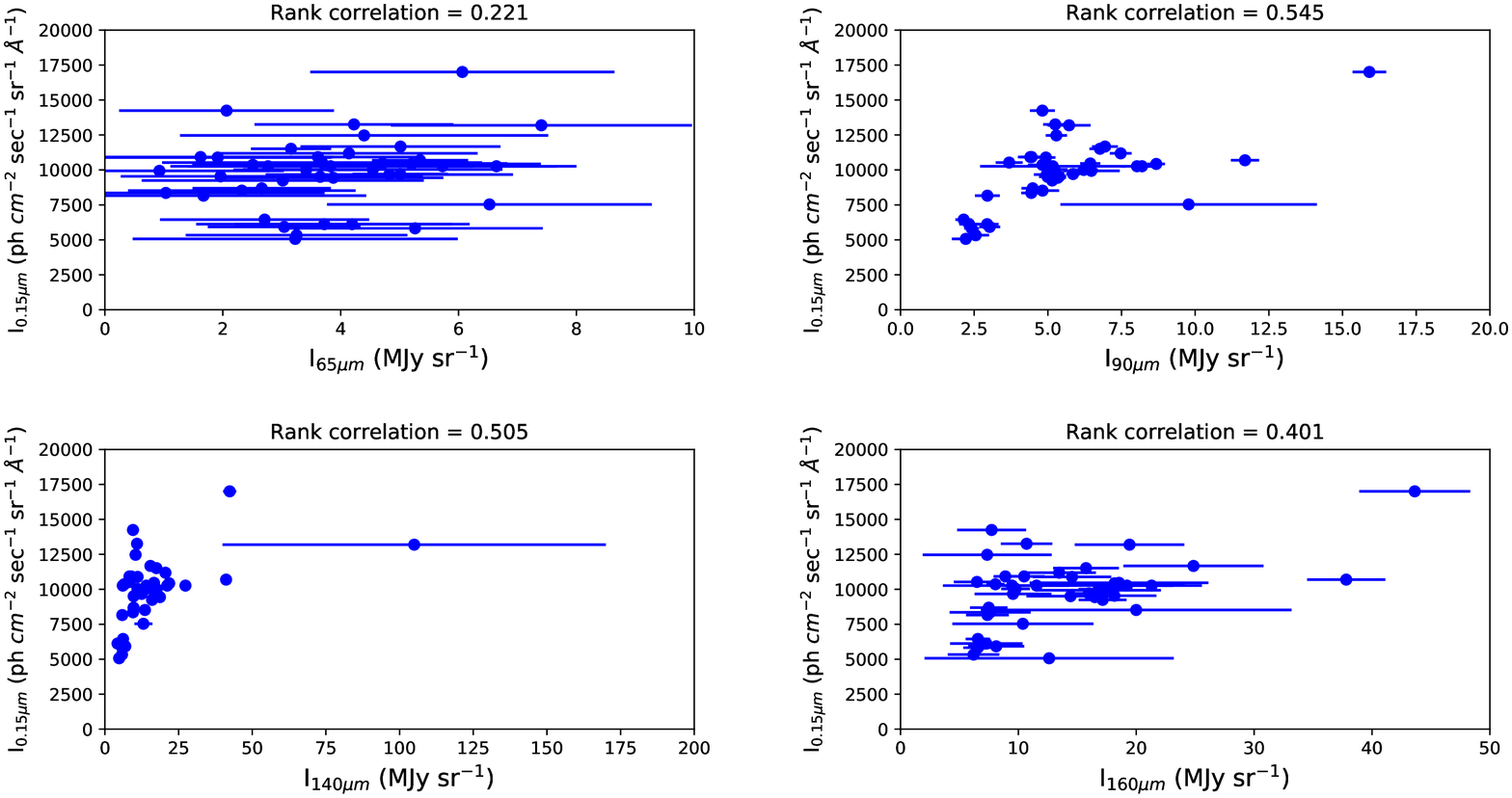}
	\caption{The FUV--IR correlations plotted for the 42 locations.}
    \label{corr_figs}
\end{figure*}

We see from our correlation studies in Table \ref{corr_table} and Figure \ref{corr_figs} that the FUV--FIR rank correlation is almost similar for emission at 90 $\mu$m and 140 $\mu$m with a slightly weaker correlation coefficient at 140 $\mu$m. The correlation is better than what is seen at 65 $\mu$m for all longer wavelengths. This shows that the dust species which shows emission at both wavelength bands around 100 $\mu$m is from similar cold environments [\cite{Seon+2011}; \cite{Seon+2011_new}; \cite{Hamden+2013}]. The emission beyond 100 $\mu$m is from colder and larger grains and hence the weaker correlation as we move to longer wavelengths. Now, emission at 65 $\mu$m is associated with star-forming regions [\cite{Onaka+2003}]. Therefore, the dust grains are prone to destruction in the presence of high UV radiation fields [\cite{Madden+2000}; \cite{Galliano+2005}] and hence the observed low value of correlation coefficient. The overall correlation trend indicates that the dust contributing to the scattering in our locations away from the centre of the Orion nebula is associated with colder environments as compared to the centre which has the Trapezium star cluster. This is in agreement with the previous work done by \cite{Shalima+2006} where it is seen that dust in the neutral HI sheet is responsible for the scattering and not dust in the HII region. So this thin sheet of dust may be extending even for our locations but at a slightly different distance.\\

\begin{figure*}
	\includegraphics[width=15cm]{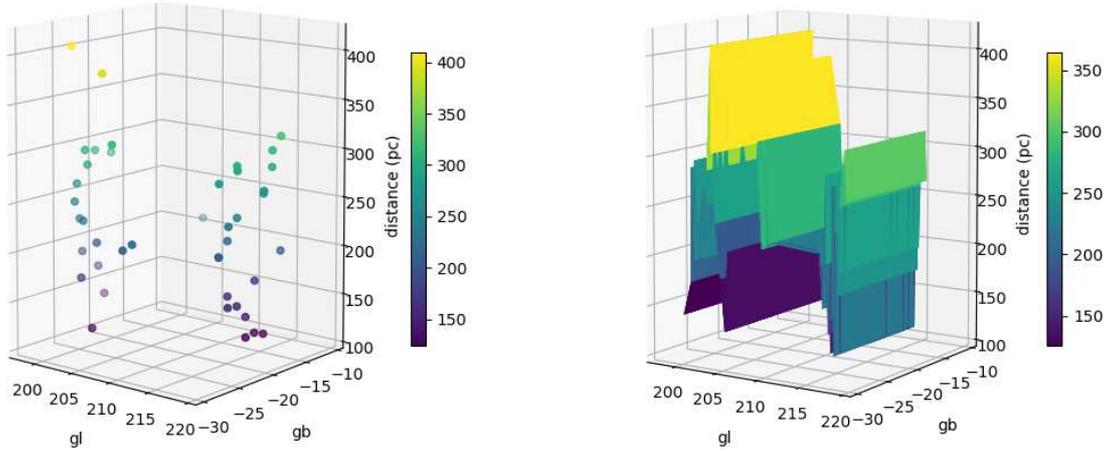}
	\caption{The distances to our dust locations shown as a 3D distribution on a scatter (left) and surface (right) plot.}
    \label{dust_dist}
\end{figure*}

Our model gives the outputs separately for each combination of $\alpha$, g and distance of dust cloud. We have done our calculations for a dust sheet 0.1 pc thick with input parameters as discussed in section 2.3. We compare our model results with the observed GALEX FUV values from Table \ref{galex_data} by placing the dust at various distances ranging from 100 pc to 450 pc. We get the best fit $\alpha$, g and distance values for the different dust locations from our model and as shown in Table \ref{agd_values}. We see higher deviation between the model and observed values as we move the dust farther away towards 450 pc from the distances specified in Table \ref{agd_values}. Since the dust grains are forward scattering, it is expected that the dust responsible for the scattering should be in front of the stars which is evident from our distance calculations as shown in Figure \ref{dust_dist}. In order to check this argument, we have removed all the dust behind the stars to see the contribution from foreground dust. We find that the change in output varies from 0.4--16$\%$ which means that atleast 84$\%$ of the light scattering is from foreground dust. So we can say that the background dust contribution is negligible.\\

We have tried to check the results by grouping our locations into four regions according to their position in Figure \ref{our_locations}. Group 1: single location on top of the 14 stars at the centre, Group 2: 15 locations to the left, Group 3: 4 locations to the bottom, Group 4: 22 locations to the right of the central region. We see that there is a consistency for all the groups irrespective of the distance and location and although individual locations have varying $\alpha$ and g with distance (Table \ref{agd_values}), each group has the same median values for these parameters, i.e. $\alpha$ = 0.7, g = 0.6 at 1539 \AA. This is an increase in the value of the albedo from those observed by \cite{Shalima+2006} at lower wavelengths for their sample. This is also higher than the theoretically predicated value of $\alpha$ = 0.4 by \cite{Draine+2003} at similar wavelengths for average Milky Way dust with R${_v}$ = 5.5. However, this is the median value for our locations and we do see a few individual locations in Table \ref{agd_values} with albedo values matching with the predictions of \cite{Draine+2003} model.\\

\begin{table*}
\centering
\caption{Our best fit parameters at different locations.}
	\label{agd_values}
	\begin{tabular}{ccccc} 
		\hline
		l  & b & $\alpha$ & g & d \\
		($^0$) & ($^0$) &  &  & (pc) \\
		\hline
215.9858 &  -21.03678  & 0.7 & 0.6 & 221.7 \\
216.3477 &  -20.05170 &  0.4 & 0.7 & 243.3 \\
216.6843 &  -21.82404  & 0.4 & 0.6 & 168.9 \\
216.3610 &  -22.73478 & 0.7 & 0.6 & 209.2 \\
217.0416&  -20.85384  & 0.9 & 0.6 & 297.6 \\
217.3502&  -19.92887  & 0.7 & 0.6 & 145.1 \\
217.6604 &  -18.98852  & 0.6 & 0.6 & 127.3 \\
203.7195 &  -26.30023  & 0.6 & 0.8 & 212.3 \\
217.4199 &  -22.61161  & 0.6 & 0.4 & 160.5 \\
217.7529 &  -21.67958  & 0.7 & 0.6 & 160.5 \\
217.9658 &  -18.02852 & 0.5 & 0.9 & 124.5 \\
217.0605 &  -23.54354  & 0.7 & 0.6 & 212.3 \\
218.0538 &  -20.73285 & 0.8 & 0.0 & 127.5 \\
203.1845 &  -26.43842  & 0.7 & 0.6 & 124.5 \\
218.3820 &  -19.80989  & 0.7 & 0.6 & 183.1 \\
200.9855 &  -24.76113 & 0.7 & 0.6 & 284.9 \\
199.4499 &  -22.13578  & 0.7 & 0.6 & 293.2 \\
199.0548 &  -19.42800 & 0.7 & 0.6 & 285.7 \\
201.5206 &  -25.63027 & 0.7 & 0.6 & 302.1 \\
218.2467 &  -17.08630  & 0.4 & 0.2 & 291.5 \\
218.6914 &  -18.85640  & 0.7 & 0.7 & 268.8 \\
198.9741 &  -21.24889 & 0.3 & 0.1 & 170.3 \\
209.4986 &  -28.72108 & 0.3 & 0.2 & 218.3 \\
210.6291 &  -28.62830 & 0.7 & 0.8 & 225.7 \\
202.0663 &  -26.48617  & 0.7 & 0.6 & 232.5 \\
218.4705 &  -22.45449  & 0.7 & 0.6 & 297.6 \\
218.1448 &  -23.38423  & 0.7 & 0.6 & 243.9 \\
217.7865 &  -24.31932  & 0.6 & 0.7 & 287.3 \\
218.9770 &  -17.90198 & 0.9 & 0.1 & 310.5 \\
218.5336 &  -16.11330 & 0.2 & 0.2 & 205.7 \\
202.6347 &  -27.34658  & 0.1 & 0.5 & 177.3 \\
199.3837 &  -23.90570 & 0.4 & 0.9 & 191.9 \\
199.8962 &  -24.77457 & 0.6 & 0.3 & 228.3 \\
219.4075 &  -19.65953 & 0.7 & 0.6 & 273.9 \\
200.4167 &  -25.65025 & 0.4 & 0.0 & 266.6 \\
203.5377 &  -10.56900 & 0.7 & 0.6 & 205.7 \\
198.0511 &  -19.44966 & 0.7 & 0.6 & 135.8 \\
219.2478 &  -16.95318 & 0.6 & 0.0 & 323.6 \\
200.9614 &  -26.51945 & 0.7 & 0.6 & 250.6 \\
209.0111 &  -29.63168  & 0.7 & 0.6 & 323.6 \\
207.8586 &  -29.71661  & 0.7 & 0.6 & 392.1 \\
201.5159 &  -27.37775 & 0.1 & 0.9 & 409.8 \\
		\hline
	\end{tabular}
\end{table*}


\section{Conclusions}

\begin{itemize}

\item We find the dust grains contributing to the extinction in our locations to be associated with colder environments as compared to the central Orion region (with the Trapezium star cluster) which agrees with the observations made by \cite{Shalima+2006}, where Orion's veil is seen to be responsible for the scattering and not the HII region dust. We also see better correlation values at longer wavelengths indicating the origin of the emission to be from larger sized dust grains. This is in agreement to the findings of a lack of small dust particles by \cite{Beitia-Antero+2017} as confirmed by the decrease in the strength of the 2175 \AA \hspace{0.1cm} feature. The reason might be due to possible destruction of PAHs or photo evaporation of the small dust grains in sites of heavy irradiation as evident from the low correlation values seen at shorter wavelengths (Table \ref{corr_table}).\\

\item  The FUV light scattering observed by us is predominantly from the foreground dust which provides for atleast 84$\%$ of the scattered radiation. \cite{Schlafly+2015} have found a 14$^0$ circular ring of dust engulfing the star-forming regions as well as dust clouds in Orion and they have estimated all the material to be lying between 400--550 pc giving a 150 pc depth and 100 pc width to the ring. Almost all of our dust locations lie within the 100--400 pc range (Table \ref{agd_values}) and hence we are less concerned with the negligible background dust contaminating our observed results.\\

\item We find the median values for our model parameters to be $\alpha$ = 0.7, g = 0.6, which are same for all the four groups in Figure \ref{our_locations} irrespective of distance or location. Although these values are higher than those predicted by \cite{Draine+2003}, we do find consistency in results of few individual locations (Table \ref{agd_values}). This might be attributed to the presence of larger sized dust grains (as seen from the correlation studies) at our locations leading to high extinction values.\\

\item From our obtained $\alpha$ and g values, we can conclude that the the dust grain properties at our observed locations are significantly different from those close to the centre which was expected. This is supported by observations made by \cite{Scandariato+2011} showing that the central Orion molecular complex, towards the direction of the Trapezium cluster, accounts for the largest amount of extinction with a decrease in the extinction towards the edge of the nebula. This indicates that the thin layer of neutral HI doesn't extend as far as our observed locations.

\end{itemize}

\section*{Acknowledgements}

We thank an anonymous referee for useful comments and suggestions which have helped in improving this paper. This work is based on observations with AKARI, a JAXA project with the participation of ESA. Some of the data presented in this paper was obtained from the Mikulski Archive for Space Telescopes (MAST). STScI is operated by the Association of Universities for Research in Astronomy, Inc., under NASA contract NAS5-26555. Support for MAST for non-HST data is provided by the NASA Office of Space Science via grant NNX09AF08G and by other grants and contracts. This research has made use of the SIMBAD database, operated at CDS, Strasbourg, France. RG would like to thank the IUCAA associateship programme for their support and hospitality.




\bibliographystyle{mnras}
\bibliography{mybibfile} 




\begin{table*}
	\centering
	\caption{Details of GALEX observations}
	\label{galex_data}
	\begin{tabular}{lcccccr} 
		\hline
		Survey & l  & b & Angular distance  & Exp. time  & FUV intensity (1539 \AA)   \\
		& ($^0$) & ($^0$) & (degrees) & (sec) & (ph $cm^{-2}$ sec$^{-1}$ sr$^{-1}$ \AA$^{-1}$) \\
		\hline
		AIS & 215.9858 &  -21.03678 & 6.96443 & 216 & 13192.089   $\pm$  43.425    \\
		AIS & 216.3477 &  -20.05170    & 7.14608 & 247 & 11670.878  $\pm$   38.819   \\
		AIS & 216.6843 &  -21.82404    & 7.80018 & 104 & 10691.843  $\pm$   55.926    \\
		AIS & 216.3610 &  -22.73478     & 7.84751 & 213 & 9540.414  $\pm$   14.409    \\
		AIS & 217.0416 &  -20.85384    & 7.89419 & 390.15 & 10889.231  $\pm$  135.868    \\
		AIS & 217.3502 &  -19.92887     & 8.07774 & 198 & 9935.189  $\pm$   61.848    \\
		AIS & 217.6604 &  -18.98852     & 8.37539 & 190 & 9244.332  $\pm$   17.995    \\
		GI & 203.7195 &  -26.30023     & 8.41982 & 1654.15 & 7533.637  $\pm$   27.338    \\
		AIS & 217.4199 &  -22.61161     & 8.70199 & 186 & 9441.720  $\pm$   29.970    \\
		AIS & 217.7529 &  -21.67958    & 8.71578 & 211.05 & 10264.169  $\pm$    4.507    \\
		AIS & 217.9658 &  -18.02852    & 8.77530 & 199 & 10691.843  $\pm$   10.494    \\
		AIS & 217.0605 &  -23.54354    & 8.78004 & 191.1 & 10428.659  $\pm$   59.874    \\
		AIS & 218.0538 &  -20.73285    & 8.81077 & 199 & 11514.293  $\pm$   30.233    \\
AIS & 203.1845 &  -26.43842     & 8.81313 & 169 & 6448.004  $\pm$   22.732    \\
AIS & 218.3820 &  -19.80989    & 9.04263 & 136 & 10264.169  $\pm$   76.981    \\
AIS & 200.9855 &  -24.76113    & 9.06785 & 176 & 10922.129  $\pm$   22.765    \\
AIS & 199.4499 &  -22.13578    & 9.19124 & 112 & 10527.353  $\pm$  107.905    \\
AIS & 199.0548  & -19.42800    & 9.19351 & 181.05 & 13257.885  $\pm$   68.427    \\
AIS & 201.5206 &  -25.63027     & 9.23280 & 188 & 8158.699  $\pm$   15.560    \\
AIS & 218.2467 &  -17.08630    & 9.24187 & 194 & 10461.557  $\pm$   26.219    \\
AIS & 218.6914 &  -18.85640    & 9.35742 & 217 & 10000.985  $\pm$   19.640   \\
AIS & 198.9741 &  -21.24889    & 9.41884 & 110 & 10922.129  $\pm$  196.400    \\
AIS & 209.4986 &  -28.72108     & 9.44258 & 283 & 9507.516  $\pm$   45.728    \\
AIS & 210.6291 &  -28.62830     & 9.47636 & 168 & 9672.006  $\pm$   38.819    \\
AIS & 202.0663 &  -26.48617     & 9.48874 & 184.05 & 5329.472  $\pm$   34.213    \\
AIS & 218.4705 &  -22.45449     & 9.56769 & 265 & 8356.086   $\pm$  52.636    \\
AIS & 218.1448 &  -23.38423     & 9.61104 & 204.05 & 8520.576  $\pm$   13.619    \\
AIS & 217.7865 &  -24.31932     & 9.73240 & 201.05 & 8685.066  $\pm$    6.941    \\
AIS & 218.9770 &  -17.90198    & 9.74467 & 213.1 & 10264.169  $\pm$   60.532  \\
AIS & 218.5336 &  -16.11330    & 9.80239 & 197 & 10264.169  $\pm$   38.819    \\
AIS & 202.6347 &  -27.34658     & 9.83424 & 189 & 6119.024  $\pm$   44.083    \\
AIS & 199.3837 &  -23.90570    & 9.88809 & 174 & 10362.863  $\pm$   66.124    \\
AIS & 199.8962 &  -24.77457    & 9.89899 & 184.05 & 12468.334  $\pm$  246.734    \\
AIS & 219.4075  & -19.65953     & 10.00526        & 220 & 9704.904  $\pm$   55.597    \\
AIS & 200.4167 &  -25.65025     & 10.01125        & 185 & 5921.636  $\pm$   69.414    \\
GI & 203.5377 &  -10.56900    & 10.10049        & 2446.25 & 17008.255 $\pm$   103.957    \\
AIS & 198.0511 &  -19.44966    & 10.13952        & 192 & 14244.825  $\pm$  106.589    \\
AIS & 219.2478 &  -16.95318    & 10.20073        & 273 & 11185.313  $\pm$   34.542    \\
AIS & 200.9614 &  -26.51945     & 10.20460        & 187 & 5822.942  $\pm$    2.957    \\
AIS & 209.0111 &  -29.63168    & 10.33346        & 288.05 & 10033.883  $\pm$  248.708    \\
AIS & 207.8586 &  -29.71661     & 10.45161        & 296 & 6119.024  $\pm$   28.851    \\
AIS & 201.5159 &  -27.37775     & 10.48114        & 188 & 5066.288  $\pm$   57.242   \\
		\hline
	\end{tabular}
\end{table*}


\begin{table*}
	\centering
	\caption{Details of AKARI observations}
	\label{akari_data}
	\begin{tabular}{lcccccc} 
		\hline
		l  & b & $I_{65 \mu m}$ & $I_{90 \mu m}$ & $I_{140 \mu m}$ & $I_{160 \mu m}$ \\
		($^0$) & ($^0$) & (MJy sr$^{-1}$) & (MJy sr$^{-1}$) & (MJy sr$^{-1}$) & (MJy sr$^{-1}$) \\
		\hline
		 215.9858 &  -21.03678 & 7.406  $\pm$ 2.554 &  5.720  $\pm$ 0.728 & 104.965  $\pm$ 65.045 & 19.430  $\pm$  4.645 \\
		 216.3477 &  -20.05170    & 5.016  $\pm$  1.697 &  6.939  $\pm$ 0.437 &  15.441  $\pm$  0.788 & 24.855  $\pm$  5.945 \\
		 216.6843 &  -21.82404    & 1.679  $\pm$  1.179  & 5.685  $\pm$ 0.494 &  16.532  $\pm$  0.979 & 15.916  $\pm$  0.651 \\
		 216.3610 &  -22.73478     & 1.962  $\pm$  1.694 &  5.385 $\pm$  0.233 &  17.261  $\pm$  1.726 & 18.142  $\pm$  3.573 \\
		 217.0416 &  -20.85384    & 1.914  $\pm$  1.931 &  4.932  $\pm$ 0.324 &  11.089  $\pm$  0.763 & 14.551   $\pm$ 3.319   \\
		 217.3502 &  -19.92887     & 0.925  $\pm$  2.111 &  6.476  $\pm$ 0.965 &  13.311  $\pm$  1.801 & 16.783  $\pm$  5.309 \\
		 217.6604 &  -18.98852     & 3.018   $\pm$ 2.392  & 5.147  $\pm$ 0.247 &  16.001  $\pm$  0.735 & 17.148  $\pm$  2.031  \\
		 203.7195 &  -26.30023     & 6.525  $\pm$  2.755 &  9.778  $\pm$ 4.350 &  13.059  $\pm$  3.042 & 10.379  $\pm$  5.982  \\
		 217.4199 &  -22.61161     & 3.875  $\pm$  1.866 &  5.337  $\pm$ 0.273  & 18.672  $\pm$  1.157 & 16.480  $\pm$  0.652 \\
	 217.7529 &  -21.67958    & 6.643  $\pm$  1.358 &  5.168 $\pm$  2.467  &  6.064  $\pm$  0.367 &  9.464   $\pm$ 5.861   \\
		 217.9658 &  -18.02852    & 5.347  $\pm$  0.815 & 11.691  $\pm$ 0.478 &  41.122  $\pm$  1.037 & 37.812  $\pm$  3.314 \\
		 217.0605 &  -23.54354    & 5.202   $\pm$ 2.196 &  8.678  $\pm$ 0.295 &  21.808  $\pm$  0.946 & 18.154   $\pm$ 5.948    \\
		 218.0538 &  -20.73285    & 3.158   $\pm$ 0.680  & 6.768  $\pm$ 0.355  & 17.483  $\pm$  0.956 & 15.739   $\pm$ 2.787 \\
 203.1845 &  -26.43842     & 2.710  $\pm$  1.774 &  2.143  $\pm$ 0.280  &  6.151  $\pm$  0.280 &  6.563 $\pm$   1.035 \\
 218.3820 &  -19.80989    & 3.828  $\pm$  2.425 &  4.938  $\pm$ 0.319  & 14.072   $\pm$ 1.020 & 11.541 $\pm$   2.761 \\
 200.9855 &  -24.76113    & 3.615   $\pm$ 1.771 &  4.406  $\pm$ 0.330  &  8.334  $\pm$  0.391 &  8.891 $\pm$   0.998 \\
 199.4499 &  -22.13578    & 3.685  $\pm$  2.715 &  3.687  $\pm$ 0.458 &   8.420  $\pm$  0.097  & 6.484 $\pm$   1.966 \\
 199.0548  & -19.42800    & 4.226   $\pm$ 1.688  & 5.250  $\pm$ 0.411  & 10.893  $\pm$  1.502 & 10.694 $\pm$   2.178 \\
 201.5206 &  -25.63027     & 1.672  $\pm$  2.761 &  2.947 $\pm$  0.421 &   5.884  $\pm$  0.491 &  7.373  $\pm$  1.816 \\
 218.2467 &  -17.08630    & 4.712  $\pm$  2.117 &  6.440 $\pm$  0.341 &  16.650  $\pm$  1.496 & 18.540 $\pm$   7.587 \\
 218.6914 &  -18.85640    & 3.415  $\pm$  1.235 &  6.217  $\pm$ 0.482 &  17.582   $\pm$ 0.478 & 17.372 $\pm$   2.266 \\
 198.9741 &  -21.24889    & 1.623   $\pm$ 2.254 &  4.448  $\pm$ 0.463  &  8.919  $\pm$  0.600 & 10.488 $\pm$   1.734 \\
 209.4986 &  -28.72108     & 3.653  $\pm$  1.743 &  4.990  $\pm$ 0.287  &  9.750  $\pm$  0.624 & 14.427 $\pm$   3.743 \\
 210.6291 &  -28.62830     & 4.826   $\pm$ 2.098  & 4.961  $\pm$ 0.434  & 12.482   $\pm$ 1.063  & 9.541  $\pm$  3.260 \\
 202.0663 &  -26.48617     & 3.252  $\pm$  1.880 &  2.550  $\pm$ 0.455  &  5.799  $\pm$  0.815 &  6.197 $\pm$   2.183 \\
 218.4705 &  -22.45449     & 1.034   $\pm$ 2.694 &  4.438  $\pm$ 0.342  &  9.545   $\pm$ 0.448 &  7.608 $\pm$   3.438 \\
 218.1448 &  -23.38423     & 2.323  $\pm$  1.931  & 4.819  $\pm$ 0.562 &  13.581   $\pm$ 0.697 & 19.994 $\pm$  13.181 \\
 217.7865 &  -24.31932     & 2.661  $\pm$  1.174  & 4.486  $\pm$ 0.378  &  9.728   $\pm$ 0.433 &  7.484 $\pm$   1.591 \\
 218.9770 &  -17.90198    & 2.768  $\pm$  1.656 &  8.024  $\pm$ 0.545  & 21.128  $\pm$  1.697 & 21.301 $\pm$   4.265 \\
 218.5336 &  -16.11330    & 5.729   $\pm$ 1.934 &  8.196  $\pm$ 0.364 &  27.330  $\pm$  1.727 & 19.215  $\pm$  3.904 \\
 202.6347 &  -27.34658     & 3.723  $\pm$  2.174  & 2.328  $\pm$ 0.335  &  4.282  $\pm$  0.399 &  7.278 $\pm$   3.074\\
 199.3837 &  -23.90570    & 2.510  $\pm$  1.028  & 4.815 $\pm$  0.487  &  6.580   $\pm$ 0.918 &  8.053 $\pm$   0.769 \\
 199.8962 &  -24.77457    & 4.398   $\pm$ 3.124 &  5.288  $\pm$ 0.361 &  10.406  $\pm$  0.718  & 7.349 $\pm$   5.469 \\
 219.4075  & -19.65953     & 5.010  $\pm$  0.809 &  5.855  $\pm$ 0.233 &  16.949  $\pm$  0.970 & 15.977 $\pm$   1.486 \\
 200.4167 &  -25.65025     & 3.042   $\pm$ 1.296 &  3.026  $\pm$ 0.355  &  6.915  $\pm$  0.239 &  8.119  $\pm$  2.381      \\
 203.5377 &  -10.56900    & 6.065   $\pm$ 2.582 & 15.911 $\pm$  0.571 &  42.355  $\pm$  2.233 & 43.627 $\pm$   4.707 \\
 198.0511 &  -19.44966    & 2.064  $\pm$  1.822 &  4.814  $\pm$ 0.424  &  9.527  $\pm$  0.924 &  7.729 $\pm$   2.922 \\
 219.2478 &  -16.95318    & 4.141  $\pm$  2.184 &  7.472  $\pm$ 0.368  & 20.543   $\pm$ 1.221 & 13.462 $\pm$   3.116  \\
 200.9614 &  -26.51945     & 5.263   $\pm$ 2.166 &  2.411  $\pm$ 0.250  &  5.872   $\pm$ 0.450 &  6.565 $\pm$   1.233 \\
 209.0111 &  -29.63168    & 4.545  $\pm$  2.201  & 5.229  $\pm$ 0.436 &  11.048  $\pm$  0.372 &  9.767  $\pm$  1.216 \\
 207.8586 &  -29.71661     & 4.195  $\pm$  1.992 &  2.940  $\pm$ 0.386 &   6.128  $\pm$  1.154 &  7.132 $\pm$   0.605 \\
 201.5159 &  -27.37775     & 3.229  $\pm$  2.758  & 2.214  $\pm$ 0.472  &  4.830  $\pm$ 0.380 & 12.611 $\pm$  10.570 \\
		\hline
	\end{tabular}
\end{table*}



\bsp	
\label{lastpage}
\end{document}